\documentclass[%
 reprint,
 amsmath,amssymb,
 aps,
]{revtex4-2}

\usepackage{graphicx}
\usepackage{dcolumn}
\usepackage{bm}
\usepackage[dvipsnames]{xcolor}
\usepackage{orcidlink}
\usepackage{aas_macros}

\newcommand{\vect}[1]{\boldsymbol{#1}}
\newcommand\norm[1]{\left\lVert#1\right\rVert}

\begin{document}

\preprint{APS/123-QED}


\title{Fast Baryonic Field Painting for Sunyaev-Zel'dovich Analyses: Transfer Function vs. Hybrid Effective Field Theory}
\author{ R. Henry Liu,$^{1,2,3}$ \orcidlink{0000-0001-6079-5683}}
\email{rh\_liu@berkeley.edu}
\author{Boryana Hadzhiyska$^{2,3,4}$}
\author{Simone Ferraro$^{2,3}$}
\author{Sownak Bose$^{5}$}
\author{C\'esar Hern\'andez-Aguayo$^{6}$}

\affiliation{
$^{1}$Department of Physics, University of California, Berkeley, CA 94720, USA \\
$^{2}$Lawrence Berkeley National Laboratory, 1 Cyclotron Road, Berkeley, CA 94720, USA \\
$^{3}$Berkeley Center for Cosmological Physics, Department of Physics, University of California, Berkeley, CA 94720, USA \\
$^{4}$Miller Institute for Basic Research in Science, University of California, Berkeley, CA, 94720, USA \\
$^{5}$Institute for Computational Cosmology, Department of Physics, Durham University, South Road, Durham, DH1 3LE, UK \\
$^{6}$Max-Planck-Institut f\"ur Astrophysik, Karl-Schwarzschild-Str. 1, D-85748, Garching, Germany \\
$^{7}$Excellence Cluster ORIGINS, Boltzmannstrasse 2, D-85748 Garching, Germany
}


\date{Accepted XXX. Received YYY; in original form ZZZ}

\date{\today}

\begin{abstract}
We present two approaches for ``painting'' baryonic properties relevant to the Sunyaev-Zel'dovich (SZ) effect -- optical depth and Compton-$y$ -- onto 3-dimensional $N$-body simulations, using the MillenniumTNG suite as a benchmark. The goal of these methods is to produce fast and accurate reconstruction methods to aid future analyses of baryonic feedback using the SZ effect. The first approach employs a Gaussian Process emulator to model the SZ quantities via a transfer function, while the second utilizes Hybrid Effective Field Theory (HEFT) to reproduce these quantities within the simulation.
Our analysis involves comparing both methods to the true MillenniumTNG optical depth and Compton-$y$ fields using several metrics, including the cross-correlation coefficient, power spectrum, and power spectrum error. Additionally, we assess how well the reconstructed fields correlate with dark matter haloes across various mass thresholds.
The results indicate that the transfer function method yields more accurate reconstructions for fields with initially high correlations ($r \approx 1$), such as between the optical depth and dark matter fields. Conversely, the HEFT-based approach proves more effective in enhancing correlations for fields with weaker initial correlations ($r \sim 0.5$), such as between the Compton-$y$ and dark matter fields. Lastly, we discuss extensions of our methods to improve the reconstruction performance at the field level.
\end{abstract}

\maketitle



\section{Introduction}
\label{sec:intro}

The distribution of baryonic content in the universe is an important component for validating cosmological models.
Baryons make up roughly 5\% of the Universe's energy density (or about 16\% of the total mass), as confirmed by measurements of the Cosmic Microwave Background (CMB) \citep{Planck2018vyg}. However, much of the universe's baryon content is still not well understood. Observations at late times only account for a small fraction of the predicted cosmological abundance of baryons \citep{Fukugita2004}. 

One important way to map the baryonic matter in the universe is through measurements of the Sunyaev-Zel'dovich (SZ) effects. CMB photons interact with free electrons in galaxies and clusters at late time, which leaves an imprint on the CMB known as the SZ effect(s) \citep{Sunyaev1970, Sunyaev1972, Sunyaev1980a, Sunyaev1980b, Carlstrom2002}. There are two main types of SZ effects, known as the thermal SZ (tSZ) effect and the kinematic SZ (kSZ) effect. 
The tSZ effect is caused by the scattering of CMB photons off electrons in hot gas surrounding haloes. It is thus proportional to the electron number density and temperature of the gas and is a probe of the thermal pressure of ionized gas around haloes.
Meanwhile, the kSZ effect is caused by the scattering of CMB photons off free electrons experiencing bulk motion relative to the CMB rest frame. This effect depends on the product between the integrated electron density along the line of sight (otherwise known as the optical depth, or $\tau$) and the peculiar velocity of individual haloes. The kSZ effect thus can act as a probe of the gas distribution on the outskirts of galaxies and haloes. For a more detailed review of the SZ effects and their physical properties, see \citep{Birkinshaw1999, Carlstrom2002, Mroczkowski2019, Bianchini2025ksz}.

Numerical simulations serve as a key tool for enhancing our understanding of the SZ effects and baryon distribution of the universe. Full-physics hydrodynamic simulations incorporate both dark matter and gas physics \cite{Jenkins1998, Teyssier2002, Di_Matteo2005}. These simulations are able to provide detailed representations of the composition and distribution of matter in the universe. Numerical simulations also make it possible to re-create mock observables such as the SZ effect \cite{Springel2005, Springel2006, Vogelsberger2020}. However, the accuracy of these simulations comes at the expense of long computation times. Large-scale full-physics hydrodynamic simulations, such as IllustrisTNG \cite{Nelson2019_IllustrisTNG}, can take millions of CPU hours to run on modern supercomputers. Furthermore, the relative scarcity of these complex, high-resolution simulations makes it difficult to form a full statistical set that can encompass varying cosmologies for inference purposes.

In contrast, dark-matter-only $N$-body simulations are significantly faster to produce and can model dark matter clustering with very high accuracy.
There exists a number of high-resolution $N$-body simulation suites with broad cosmologies, including Quijote \cite{Villaescusa-Navarro2020_Quijote}, Aemulus \cite{DeRose2019_AEMULUS}, and AbacusSummit \cite{Maksimova2021_ABACUSSUMMIT} among many others.
Furthermore, particle mesh based $N$-body simulation codes such as FastPM \cite{Feng2016_FastPM, Bayer2021_FastPM} are able to produce approximate $N$-body simulations extremely quickly, making it straightforward to generate a large number of simulations with varying cosmological parameters for inference. 
The current capabilities of generating fast $N$-body simulation suites make it particularly interesting to consider methods that could recreate the statistics of hydrodynamic simulations using $N$-body as a basis. 

As a result, ``baryon pasting'' or ``baryonification'' methods, meant to reproduce baryonic effects on $N$-body simulations, is an important task in computational cosmology.
Methods for reconstructing gas and SZ properties have been developed in both 2D map space and 3D simulation space \cite{Sehgal2010, Stein2020, Osato2023, Anbajagane2024}. Furthermore, recent advances in machine learning (ML) and the increased availability of computational power have also led to the emergence of a number of ML-based field reconstruction techniques \cite{Dai2021, Dai2022_TRENF, Dai2023, Han2021, Hong2021, Villaescusa-Navarro2021a, Villaescusa-Navarro2021b}, including some specifically aimed at reconstructing gas and SZ properties \cite{Thiele2020, Thiele2022}.
Broadly, these reconstruction methods fall into two main categories: halo-based and field-level approaches. Halo-based reconstruction methods rely on modelling the Halo Occupation Distribution (HOD) \cite{Berlind2002}, halo assembly models \cite{Behroozi2019}, or the use of other methods (such as ML \cite{Chadayammuri2023}) to model gas and feedback mechanisms centred around halo clusters for baryon pasting. In contrast, field-level methods reconstruct the full distribution of gas and baryonic properties across the entire field, offering a more comprehensive approach to baryon pasting.

Field-level reconstructions have the benefit of being able to model and reconstruct gas and SZ effects away from haloes when compared to halo-based methods. However, such methods are more expensive and complex than halo-based reconstruction techniques, especially on larger scales. In particular, some ML-based field-level reconstruction techniques have been limited to smaller box sizes or reconstructions on 2D maps rather than 3D simulation boxes (for example, \cite{Dai2022_TRENF}). Although advanced training sets of cosmological simulations, such as the CAMELS suite \cite{Villaescusa-Navarro2021_CAMELS, Villaescusa-Navarro2023_CAMELS, Ni2023_Astrid_CAMELS} now exist, the limited volumes of such simulation sets (CAMELS has a $25\ h^{-1}$ Mpc box size, for example) make training ML-based methods for large scale field-level baryon pasting models a difficult challenge.

In this paper, we consider two computationally fast field-level baryon pasting methods on dark matter simulations and apply them to the large and high-resolution MillenniumTNG \cite{Pakmor2023_MTNG, Hernandez-Aguayo2023_MTNG, Hadzhiyska2023_MTNGa, Hadzhiyska2023_MTNGb, Bose2023_MTNG} simulation.
The first method follows the work of \cite{Sharma2024}, and considers a transfer function-based approach to reconstruct the two-point statistics of the target hydrodynamic simulation fields from dark matter $N$-body simulations. Using a Gaussian Process-based emulator, we can emulate and train the transfer function using a set of CAMELS simulations and apply the emulated transfer function to the MillenniumTNG simulation suite. The usage of emulators allows us to approximate the dark-matter to baryon transfer function when the true transfer function is not present, allowing us to apply the transfer function to dark-matter-only simulations in the future.

In contrast, the second method follows the formalism of the Hybrid Effective Field Theory (HEFT) framework as described in \cite{Modi2020, Hadzhiyska2021, Kokron2021, Zennaro2023}. This method combines the formalism of a Lagrangian Perturbation Theory (LPT) approach and the numerical precision of simulations to advect the fields forward in time from initial conditions, then match them to the targeted field in question through a fitting of bias parameters.

In this work, we compare the two methods and their abilities to reconstruct the optical depth and Compton-$y$ fields in the MillenniumTNG simulation suite, and evaluate the effectiveness of the two methods using both two-point statistic metrics and correlations of our reconstructed fields to dark matter haloes in the MTNG simulation suite.



This paper is structured as follows. In Section \ref{sec:simulations} we will discuss the suites of simulations we used. Section \ref{sec:theory} will discuss the theoretical Gaussian Processes and the Hybrid Effective Field Theory methods used to reconstruct baryons. Section \ref{sec:results} presents our reconstruction results and a number of metric statistics that we apply to the reconstructed fields. Section \ref{sec:discussion} will discuss these results and conclude the paper.

\section{Simulations}
\label{sec:simulations}

In this work, we use two sets of simulations. To train the transfer function emulators, we utilize the CAMELS set of simulations, while we test the results for both methods using the MillenniumTNG simulation.

\subsection{CAMELS}
\label{sec:CAMELS}

To train the Gaussian process emulator, we make use of the Cosmology and Astrophysics with MachinE Learning Simulations (CAMELS) suite \citep{Villaescusa-Navarro2021_CAMELS, Villaescusa-Navarro2023_CAMELS, Ni2023_Astrid_CAMELS}.
CAMELS is a set of 14,091 simulations, each with a comoving volume of $25 \;h^{-1}\mathrm{Mpc}$, evolved from $z=127$ to $z=0$ with $256^3$ dark matter and $256^3$ gas particles under initial conditions. In total, the set of simulations includes 6,163 $N$-body and 7,928 hydrodynamic simulations as of the writing of this paper.

The CAMELS simulations are categorized into a number of suites (Astrid, IllustrisTNG, SIMBA, etc.), based on the code for running the simulations and the set of astrophysical parameters and initial seeds. The IllustrisTNG suite, based on \cite{Nelson2019_IllustrisTNG}, contains 1,092 hydrodynamic simulations executed using the AREPO code \citep{Springel2010_AREPO, Weinberger2020_AREPO}, while the Astrid \citep{Ni2022_Astrid, Bird2022_Astrid} and SIMBA \citep{Dave2019_SIMBA} uses the MP-Gadget \citep{Feng2018} and GIZMO \citep{Hopkins2015} codes respectively.

Each simulation from CAMELS is characterized by six parameters: $\Omega_m$ and $\sigma_8$ to characterize its cosmology as well as four astrophysical parameters $A_{\mathrm{SN1}}, A_{\mathrm{SN2}}, A_{\mathrm{AGN1}}, A_{\mathrm{AGN2}}$. The two Supernova parameters $A_{\mathrm{SN1}}$ and $A_{\mathrm{SN2}}$ control the efficiency of supernova feedback, while the other two parameters, $A_{\mathrm{AGN1}}$ and $A_{\mathrm{AGN2}}$ parameterize the efficiency of feedback from supermassive black holes. 

We make use of the CAMELS Latin-Hypercube (LH) suite in this work. The Latin Hypercube set contains 1,000 $N$-body and 1,000 hydrodynamic simulations for each of the simulation suites described above. The astrophysical and cosmological parameters of each of the simulation sets are arranged as a Latin-Hypercube within the range
\begin{align}
    \Omega_m &\in [0.1, 0.5] \\
    \sigma_8 &\in [0.6, 1.0] \\
    A_{\mathrm{SN1}}, A_{\mathrm{AGN1}} &\in [0.25, 4.0] \\
    A_{\mathrm{SN2}}, A_{\mathrm{AGN2}} &\in [0.5, 2.0]
\end{align}
Furthermore, each simulation in the LH suite has a different value of initial random seed. For each simulation in the suite, CAMELS provides redshift snapshots for a number of redshifts: 34 redshift slices from $z$ of 6.0 to 0.0 in the case of IllustrisTNG and SIMBA, and 90 redshift slices from 15.0 to 0.0 in the case of the Astrid suite. 

It is important to note that, though the astrophysical parameters described above are common among the simulation suites, they do not represent the same physical effects across simulation suites. This is due to the suites being run using different codes with different subgrid models, and means further that one should not attempt to match the values of the astrophysical parameters across the simulation suites, or to train on one simulation suite and predict on another using the given astrophysical parameters. The cosmological parameters $\Omega_m$ and $\sigma_8$ do represent physical effects and could be matched across simulation suites. 

As the goal of this project is to compare the ability to reconstruct baryon fields on the MillenniumTNG simulation suite, we opt to train our Gaussian Process emulator using the IllustrisTNG sub-suite of the CAMELS simulation suite. As will be discussed in the next section, the MillenniumTNG simulation follows as an extension of IllustrisTNG using the same AREPO code, and training on the IllustrisTNG set ensures best performance due to the same simulation code structure. 

We note that it is also possible to train and predict using the other CAMELS simulation suites. \cite{Sharma2024} was able to show the success of predicting for the total matter field by training through the ASTRID suite in CAMELS, and similar success is expected here. As we have also shown in Section \ref{sec:results_TF}, the emulator trained on CAMELS transfer functions is able to reproduce a transfer function on the MillenniumTNG simulation, which has a much larger box size.

\subsection{MillenniumTNG}
\label{sec:MillenniumTNG}

The MillenniumTNG simulation suite consists of several hydrodynamic and $N$-body simulations with various box sizes and resolutions. 
MillenniumTNG makes use of the \textit{Planck} 2016 cosmology as described in \cite{Planck2016_XIII}, the same cosmology used in the previous IllustrisTNG simulation suite, (i.e. $\Omega_0 = 0.3089$, $\Omega_{\rm b} = 0.0486$, $\Omega_\Lambda = 0.6911$, $\sigma_8 = 0.8159$, $n_s = 0.9667$, and $h = 0.6774$). 
For a detailed description of the MillenniumTNG simulation suite, see \cite{Pakmor2023_MTNG, Hernandez-Aguayo2023_MTNG, Hadzhiyska2023_MTNGa, Hadzhiyska2023_MTNGb, Bose2023_MTNG}. 

In this work, we employ the largest full physics simulation box, alongside its dark matter only counterpart, with a comoving volume of $(500\;h^{-1} \mathrm{Mpc})^3$. The simulations contain $2\times 4320^3$ particles in the full physics 
case and $1080^3$ particles in the dark matter only counterpart,
pasted onto a $1080^3$ resolution grid using the Triangular Shaped Cloud algorithm. 

We note the similarities of the MillenniumTNG and IllustrisTNG simulations: The MillenniumTNG suite uses the same cosmological model as IllustrisTNG \citep{Weinberger2017, Pillepich2018a_IllustrisTNG, Pillepich2018b},  with a slightly lower resolution than the largest IllustrisTNG box TNG300. Furthermore, the MillenniumTNG simulations are also run the AREPO code used to generate IllustrisTNG and the representative CAMELS IllustrisTNG sets, though with modifications to fit the simulation into supercomputer memory (See \cite{Pakmor2023_MTNG} for details). In analogy to the naming convention for TNG300, the MillenniumTNG simulation is referred to as MTNG740 due to the box size of $500 \;h^{-1} \mathrm{Mpc}$.

We also make use of haloes from the MillenniumTNG suite in this work as a further validation method of our field-level reconstruction methods. Halo groups are found using the Friends-of-Friends (FoF) algorithm, with a linking length of $b = 0.2$. For further details in this process, we refer to \cite{Hernandez-Aguayo2023_MTNG}.  

\section{Theory and Methods}
\label{sec:theory}

In this section, we discuss the methodology used in replicating the properties of the baryon field, starting from the dark matter density field. 
We consider a dark matter field $\delta_{m}$ and a target baryonic effect field $\delta_{\mathrm{b}}$ (which we substitute in Section \ref{sec:results} for the optical depth $\tau$ field or the Compton-$y$ field), we attempt a few different approaches to creating an approximation to the baryonic field $\Tilde{\delta}_{\mathrm{b}}$ to reasonable accuracy.

\subsection{Transfer Function Emulator}
\label{sec:transfer_function_emulator}

One straightforward approach to reproducing the characteristics of the baryon distribution is by utilizing a transfer function \cite{Sharma2024}. For Gaussian or weakly non-Gaussian fields (such as the fields considered here, at least on large scales), the power spectrum serves as the most informative statistic. For fields with strong existing correlations, the power spectrum can be matched exactly by using a transfer function. If the correlation between the fields is large to start with, we expect most of the statistical properties to be approximately replicated.
At the field level, the transfer function is formally defined as
\begin{equation}
    \Tilde{\delta}_{\mathrm{b}}(\mathbf{k}) = T(k) \delta_m(\mathbf{k})
\end{equation}
where $\delta_m(\mathbf{k})$ represents the matter overdensity field in Fourier space while $\Tilde{\delta}_{\mathrm{b}}(\mathbf{k})$ represents the reconstructed baryon field, transformed through the transfer function. In the case of the transfer from the dark matter field to the baryon fields, this transfer function from the is given as:
\begin{equation}
\label{eq:TF}
    T(k) = \sqrt{\frac{P_{\mathrm{b}}(k)}{P_m(k)}}
\end{equation}

The use of a transfer function is able to provide a number of guarantees for the transformed field. The power spectrum of the output field is guaranteed to match the desired field while leaving the cross-correlation of the fields intact.

The cross correlation coefficient, defined as 
\begin{equation}
\label{eq:r_cc}
    r(k) = \frac{P_{\mathrm{cross}}(k)}{\sqrt{P_m(k)P_\mathrm{b}(k)}}
\end{equation}
is an important metric for describing the similarity of two fields. 
We note that applying any transfer function to a field does not change the cross correlations coefficient with other fields.

We further note that the true transfer function, as defined in Equation \ref{eq:TF}, is only known exactly when the target field's power spectrum is known (i.e. when there is already a well-defined target field $\delta_{\rm b}$). In cases where there is not a well-defined target field, we would need to model the transfer function using emulator-based methods.
For modelling, we consider the use of a Gaussian Process emulator. Gaussian Processes \citep{Rasmussen2006Gaussian} are a highly adaptable method in ML and statistics, with high effectiveness in applications such as regression, interpolation, and quantifying uncertainty. Gaussian Process methods are further very efficient after the initial training process, enabling the ability to quickly emulate transfer functions given cosmological and astrophysical parameters.

Following the methods first described in \cite{Sharma2024}, we make use of a Gaussian Process emulator to emulate the transfer function between the two fields. We split the data into 39 equally spaced $k$-bins in the range $0.36<k / [h\ \mathrm{Mpc}^{-1}] <10.00$, and model each bin using an independent Gaussian Process.

As described in Section \ref{sec:simulations}, we make use of the CAMELS IllustrisTNG suite of simulations as a training set to train the power spectrum ratios for our transfer function use. We train our conditional Gaussian Process on the six CAMELS parameters: $[\Omega_m, \sigma_8, A_{\mathrm{SN1}}, A_{\mathrm{SN2}}, A_{\mathrm{AGN1}}, A_{\mathrm{AGN2}}]$, and use the Gaussian process to model the target function of the power spectra ratio $\frac{P_{\mathrm{b}}(k)}{P_{m}(k)}$. 

As mentioned in Section \ref{sec:CAMELS}, however, the four astrophysical parameters are not consistent across the CAMELS simulation suites (i.e., they have different physical meanings in different suites). Moreover, as these parameters are specific to CAMELS, they are not present in the MillenniumTNG simulation. As a result, we will instead consider these four parameters as nuisance parameters we would need to tune, in order to demonstrate that our emulator is capable of providing the true transfer function. Using the differential evolution global optimizer from Scipy \cite{Virtanen2020_scipy}, we optimize the Gaussian process for these nuisance parameters to the transfer function emulator. This follows the the methods first described in \cite{Sharma2024}.

\subsection{Lagrangian Hybrid Effective Field Theory}
\label{sec:lagrangian_HEFT}

In contrast to directly modelling the optical depth field using the dark matter overdensity field at current time using a transfer function, in this section, we consider using a hybrid Lagrangian Perturbation Theory method to model the optical depth field using particles at initial conditions. This is justified because the symmetries of the problem are the same as the galaxy field typically modelled using perturbation theory, and on large scales, we expect baryons to be (biased) tracers of matter. Our goal is to explore this method's reach for modelling the gas.
For a general overview of perturbation theory methods, see \cite{Bernardeau2002, Desjacques2018, Dodelson2020}. This Lagrangian Perturbation Theory method has been extended to the context of Effective Theory of Large-Scale Structures \cite{Baumann2012, Carrasco2012, Senatore2015, McDonald2009}, applied to Lagrangian space \cite{Porto2014, Vlah2015, Chen2020, Chen2021}. Although this method has been described by many names in literature, such as Hybrid Lagrangian Perturbation Theory, going forward this method will be described as Hybrid Effective Field Theory (HEFT) in this work.

In the Lagrangian picture, we work with fluid elements as collision-less particles labelled by their initial Lagrangian position $\vect{q}$, with their dynamics described by a displacement field of $\mathbf{\Psi}(\vect{q})$, defined such that the Eulerian positions of each fluid element at any conformal time $\eta$ is $\vect{x}(\vect{q}, \eta) = \vect{q} + \mathbf{\Psi}(\vect{q}, \eta)$ \cite{Porto2014, Vlah2015}.

Adopting the Lagrangian EFT framework, we consider a bias expansion to the functional as first described in \cite{Matsubara2008} then expanded upon in \cite{Modi2020}. 
\begin{equation}
\label{eq:bias}
\begin{split}
        F(\vect{q}) = &1 + b_\delta \delta_L(\vect{q}) + b_{\delta^2} (\delta^2_L(\vect{q}) - \langle\delta^2_L(\vect{q})\rangle) \\ 
        &+ b_{s^2} (s_L^2(\vect{q})-\langle s_L^2\rangle ) + b_{\nabla^2} \nabla^2\delta_L(\vect{q})
\end{split}
\end{equation}
With $b_\delta, b_{\delta^2}, b_{s^2}$ and $b_{\nabla^2}$ as free bias parameters, noting that $s_L^2$ is the (squared) shear field, with $s_{ij} \equiv (\partial_i\partial_j/\partial^2 - \delta_{ij}/3)\delta_L$.

In order to get the real space Eulerian positions, we can advect this functional
\begin{equation}
\label{eq:advect}
    1 + \Tilde{\delta}_{\mathrm{b}}(\vect{x}) = \int d^3\vect{q} F(\vect{q}) \delta^D(\vect{x} - \vect{q} - \mathbf{\Psi}(\vect{q}))
\end{equation}
Where $\Tilde{\delta}_{\mathrm{b}}$ is the reconstructed baryon field, as opposed to the true baryon field $\delta_{\mathrm{b}}$. However, as first described in \cite{Modi2020}, this advection step, as well as the solution to $\mathbf{\Psi}(\vect{q})$ could be conducted through the use of simulations. As a non-perturbative method, simulations offer numerical ``exact'' solutions to field advection from initial conditions. At late times, this tracer field could be explicitly written as 
\begin{equation}
\begin{split}
    \Tilde{\delta}_{\mathrm{b}}(\vect{x}) = &\delta_m(\vect{x}) + b_\delta\mathcal{O}_\delta(\vect{x}) + b_{\delta^2}\mathcal{O}_{\delta^2}(\vect{x})  + \\
    &  b_{s^2}\mathcal{O}_{s^2}(\vect{x}) + b_{\nabla^2}\mathcal{O}_{\nabla^2\delta}(\vect{x}) 
\end{split}
\end{equation}
Where we have used the notation 
$\mathcal{O}_i$ to describe the $i$-th simulation-advected field. 

The combination of simulations with the LPT methodology allows us to separate nonlinear dynamics into two components: tracer bias in the initial Lagrangian field and the nonlinear dynamics of evolution from initial conditions. 
The former is described in Equation \ref{eq:bias}, while the latter is described in Equation \ref{eq:advect} utilizing simulations. This combination of theoretical bias expansion with exact simulation modelling is referred to as the HEFT method in literature \cite{Modi2020, Hadzhiyska2021, Kokron2021, Zennaro2023}.

The HEFT model could be used to fit and reproduce a target field given the Lagrangian initial conditions. Furthermore, following the work of \cite{Schmittfull2019, Kokron2022, Baradaran2024, Hadzhiyska2025}, we can further obtain the bias-free parameters through the use of field-level fits as described as  follows:

Given a set of bias parameters $b_i$, we can find the optimal parameters by solving for the least-squares problem by minimizing the objective function $S$ in $k$ space, defined as
\begin{equation}
\label{eq:objective}
    S = \int_{|\vect{k}|<k_\mathrm{max}} \frac{d^3k}{(2\pi)^3} \norm{\delta_{\mathrm{b}}(\vect{k}) - \delta_m(\vect{k}) - \sum_i b_i \mathcal{O}_i(\vect{k})}^2
\end{equation}
where $\mathcal{O}_i$ describes one of the advected $\delta_L, \delta_L^2, s_L^2, \nabla^2\delta_L$ fields and $b_i$ describes the corresponding bias free parameter.
The estimator solution to this least squares minimization problem is given by 
\begin{equation}
    \hat{b}_i = M_{ij}^{-1} A_j
\end{equation}
Where $A_j$ is defined as
\begin{align}
    A_j &= \langle [\mathcal{O}_j(\vect{x}) (\delta_{\mathrm{b}}(\vect{x}) - \delta_m(\vect{x}))]_{k_{\mathrm{max}}}\rangle \\
    &= \int_{|\vect{k}|<k_\mathrm{max}} \frac{d^3k}{(2\pi)^3} \mathcal{O}_j(\vect{k}) [\delta_{\mathrm{b}} - \delta_m]^*(\vect{k})
\end{align}
and $M_{ij}$ as
\begin{align}
\label{eq:Mij}
    M_{ij} &= \langle[\mathcal{O}_i(\vect{x})\mathcal{O}_j(\vect{x})]_{k_\mathrm{max}}\rangle \\
    &= \int_{|\vect{k}|<k_\mathrm{max}} \frac{d^3k}{(2\pi)^3} \mathcal{O}_i(\vect{k})\mathcal{O}_j^*(\vect{k})
\end{align}

We can also define a scale-dependent version of the bias parameters, replacing $b_i$ with $b_i(k)$, where we instead calculate the $k$ dependent bias:
\begin{equation}
    \hat{b}_i(k) = M_{ij}(k)^{-1} A_j(k)
\end{equation}
with $A_j(k)$ and $M_{ij}(k)$ defined as:
\begin{align}
    A_j(k) &= \int_{k_{\mathrm{min}} < |\vect{k}| < k_{\mathrm{max}}} \frac{d^3k}{(2\pi)^3} \mathcal{O}_j(\vect{k}) [\delta_{\mathrm{b}} - \delta_m]^*(\vect{k}) \\
    M_{ij}(k) &= \int_{k_{\mathrm{min}} < |\vect{k}| < k_{\mathrm{max}}} \frac{d^3k}{(2\pi)^3} \mathcal{O}_i(\vect{k})\mathcal{O}_j^*(\vect{k})
\end{align}
Note that in this case, rather than describing continuous $k$, we consider discrete $k$ bins, described by $k_{\mathrm{min}}$ and $k_{\mathrm{max}}$ as an approximation for the continuous bias parameter. This method, previously implemented in \cite{Kokron2022}, gives us four scale-dependent biases for the Lagrangian fields, which, together with $\delta_m$, would give us an approximation for $\delta_{\mathrm{b}}$.

This leaves us with a Lagrangian bias-dependent field as follows for describing the target field. Written explicitly, the scale-dependent field approximation is given by
\begin{equation}
\begin{split}
    \Tilde{\delta}_{\mathrm{b}}(\vect{k}) = &\delta_{m}(\vect{k}) + \hat{b}_{\delta}(k) \mathcal{O}_{\delta}(\vect{k}) + \hat{b}_{\delta^2}(k) \mathcal{O}_{\delta^2}(\vect{k})  \\ 
    &+ \hat{b}_{s^2}(k)\mathcal{O}_{s^2}(\vect{k}) + \hat{b}_{\nabla^2}(k)\mathcal{O}_{\nabla^2 \delta}(\vect{k})
\end{split}
\end{equation}
With the note that the $\delta_{m}$ dark matter field is the advected CDM + baryons field from Lagrangian initial conditions.

Lastly, we fit an additional amplitude parameter to $\Tilde{\delta}_{\mathrm{b}}$ in order to match the power spectra to $\delta_{\mathrm{b}}$. We do so using a weighted least-squares fit on the power spectra, using a single amplitude parameter $A$. Explicitly written, the $\Tilde{\delta}_{\mathrm{b}}$ field with modified amplitude is 
\begin{equation}
\label{eq:amplitude}
\begin{split}
    \Tilde{\delta}_{\mathrm{b}}(\vect{k}) = A \big(&\delta_{m}(\vect{k}) + \hat{b}_{\delta}(k) \mathcal{O}_{\delta}(\vect{k}) + \hat{b}_{\delta^2}(k) \mathcal{O}_{\delta^2}(\vect{k})  \\ 
    &+  \hat{b}_{s^2}(k)\mathcal{O}_{s^2}(\vect{k}) + \hat{b}_{\nabla^2 \delta}(k)\mathcal{O}_{\nabla^2 \delta}(\vect{k}) \big)
\end{split}
\end{equation}
Henceforth, the mode in Equation \ref{eq:amplitude} will be described as the scale-dependent HEFT, while the scale-independent HEFT replaces $b_i(k)$ in Equation \ref{eq:amplitude} with constant $b_i$.

\subsection{Extended Hybrid Effective Field Theory}
\label{sec:extended_HEFT}
From the previous section, one can see that a straightforward extension of the Lagrangian bias expansion method through adding a bias parameter to the dark matter overdensity field $\delta_{m}$. Though such a modification would no longer yield a strictly ``Lagrangian" bias expansion result, as we will see, it does yield a better performing model for approximating the target field. It is important to point out that while inspired by Lagrangian Perturbation Theory, this expansion is not used to fit cosmological or astrophysical parameters to the gas, can be seen as a ``symmetry-motivated'' fitting function to the relation between the dark matter and baryon fields, and therefore is not constrained to take the strict form implied by the equations of motion derived in the traditional perturbative expansions. We will further describe this method we call Extended HEFT, or E-HEFT for short.

In order to fit a bias expansion including $\delta_m$, we adopt a simple modification. Here, we would treat $\delta_m(\vect{k})$ in Equation \ref{eq:objective} as part of the $\mathcal{O}_i$ set of advected operators. Propagating this forward, this would mean that $A_j$ is now defined as 
\begin{align}
    A_j &= \langle [\mathcal{O}_j(\vect{x}) \delta_{\mathrm{b}}(\vect{x})]_{k_{\mathrm{max}}}\rangle \\
    &= \int_{|\vect{k}|<k_\mathrm{max}} \frac{d^3k}{(2\pi)^3} \mathcal{O}_j(\vect{k}) \delta_{\mathrm{b}}^*(\vect{k})
\end{align}
while $M_{ij}$ is unchanged as in Equation \ref{eq:Mij}. 

Similarly, in the scale dependent bias version, we can write the modified $A_j(k)$ as
\begin{equation}
    A_j(k) = \int_{k_{\mathrm{min}} < |\vect{k}| < k_{\mathrm{max}}} \frac{d^3k}{(2\pi)^3} \mathcal{O}_j(\vect{k}) \delta_{\mathrm{b}}^*(\vect{k}) 
\end{equation}
Explicitly, the extended Lagrangian HEFT method would yield an approximate $\Tilde{\delta}_{\mathrm{b}}$ as follows
\begin{equation}
\begin{split}
    \Tilde{\delta}_{\mathrm{b}}(\vect{k}) = &\hat{b}_{\delta_m}(\vect{k})\delta_{m}(\vect{k}) + \hat{b}_{\delta}(\vect{k}) \mathcal{O}_{\delta}(\vect{k}) + \hat{b}_{\delta^2}(\vect{k}) \mathcal{O}_{\delta^2}(\vect{k})  \\ 
    &+  \hat{b}_{s^2}(\vect{k})\mathcal{O}_{s^2}(\vect{k}) + \hat{b}_{\nabla^2 \delta}(\vect{k})\mathcal{O}_{\nabla^2 \delta}(\vect{k})
\end{split}
\end{equation}
with either $k$-dependent or independent biases depending on the case. We note that in notation, $\mathcal{O}_\delta$ is the advected Lagrangian first order delta field, while $\delta_m$ is the advected 1cb field, which represents the dark matter overdensity.

The E-HEFT formulation here is not the same as the amplitude modified fields 
given by Equation \ref{eq:amplitude}. In the E-HEFT case, we conduct a field-level bias fit to the $\delta_m$ field, while to obtain the amplitude modified field described in Equation \ref{eq:amplitude}, we perform a fit only at the power-spectrum level. This corrects the amplitude of the power spectrum, but leaves the phases unchanged and thus cross correlation unchanged. 

We further note that if we set the higher order bias terms to zero, the solution to the least squares minimization is $\hat{b}_{\delta_m}(k)=r(k)T(k)$. This means that the transfer function method presented previously is a special case of the lowest order E-HEFT, and we are interested in whether the additional terms in the expansion significantly improve upon first-order results.

As we will see in the next section, the E-HEFT method yields better cross-correlation results when compared to the regular HEFT methodology.
To some degree, this is expected, as an additional free parameter should improve the fit. However, the improvement is significant enough to take note of the method as a synthesis of Lagrangian and Eulerian methodology, which we will discuss in Section \ref{sec:discussion}. 



\section{Results}
\label{sec:results}



\subsection{Optical Depth Field}
\label{sec:optical_depth}



We first make use of the two methods described in the previous section to predict the optical depth field in the MillenniumTNG simulation. The optical depth $\tau$ is the physical quantity probed by the kSZ effect, among others, which induces a fractional temperature fluctuation in the CMB given by:
\begin{equation}
    \frac{\Delta T_\mathrm{kSZ}}{T_{\rm CMB}} = -\tau \left(\frac{v_{e, r}}{c}\right)
\end{equation}
Where $\tau$ is the desired optical depth along the observed line of sight, while $v_{e, r}$ is the free electron bulk line of sight velocity. The optical depth is defined as 
\begin{equation}
    \tau(z) \equiv \int \frac{d\chi}{1+z} n_e(\chi\vect{n}, z)\,\sigma_T\;.
\end{equation}
With $\chi$ as the comoving distance to redshift z, $n_e$ is the free electron density, $c$ is the speed of light, $\vect{n}$ the line-of-sight direction and $\sigma_T$ the Thompson cross section. Since our simulations are 3-dimensional, we can define an equivalent ``3D optical depth'' field in each simulation cell as given by Equation \ref{eq:tau_cell} below.
Notably, the 3D optical depth field traces the ionized component of the gas density, since $\>90\%$ of the gas is expected to be ionized at low redshift. Furthermore, the optical depth is a physically observable quantity, for example, through the kinematic Sunyaev-Zel'dovich effect \cite{Carlstrom2002}

The MillenniumTNG simulation outputs a number of quantities: gas cell mass ($m$), electron abundance ($x$), and internal energy ($\epsilon$). Assuming a primordial hydrogen mass fraction of $X_H=0.76$, we compute the volume-weighted electron number density, $n_{\rm e}$, for each gas particle $i$ as
\begin{equation}
V_i n_{{\rm e},i} = x_i m_i \frac{X_H}{m_p}
\end{equation}
where $m_p$ the proton mass. We then compute the 3D maps of the kSZ and the optical depth by binning the gas particles into a $1024^3$ cubic grid, so that the optical depth in cell $j$ is given by:
\begin{equation}
\tau_j = \sigma_T V_j^{-1} \sum_{i \in V_j} V_i n_{{\rm e},i} ,
\label{eq:tau_cell}
\end{equation}
where $V_j$ is the volume of each grid cell (of size 0.721 Mpc$^3$).
Given the definition of the optical depth field as given above, we see that for constant electron abundance $x$, the optical depth $\tau$ is proportional to the gas density field. At late times, the electron abundance $x$ is roughly constant, making the optical depth an important tracer for the underlying gas density distribution.


%

\subsubsection{Transfer function emulation}
\label{sec:results_TF}

\begin{figure}
    \centering
    \includegraphics[width=\columnwidth]{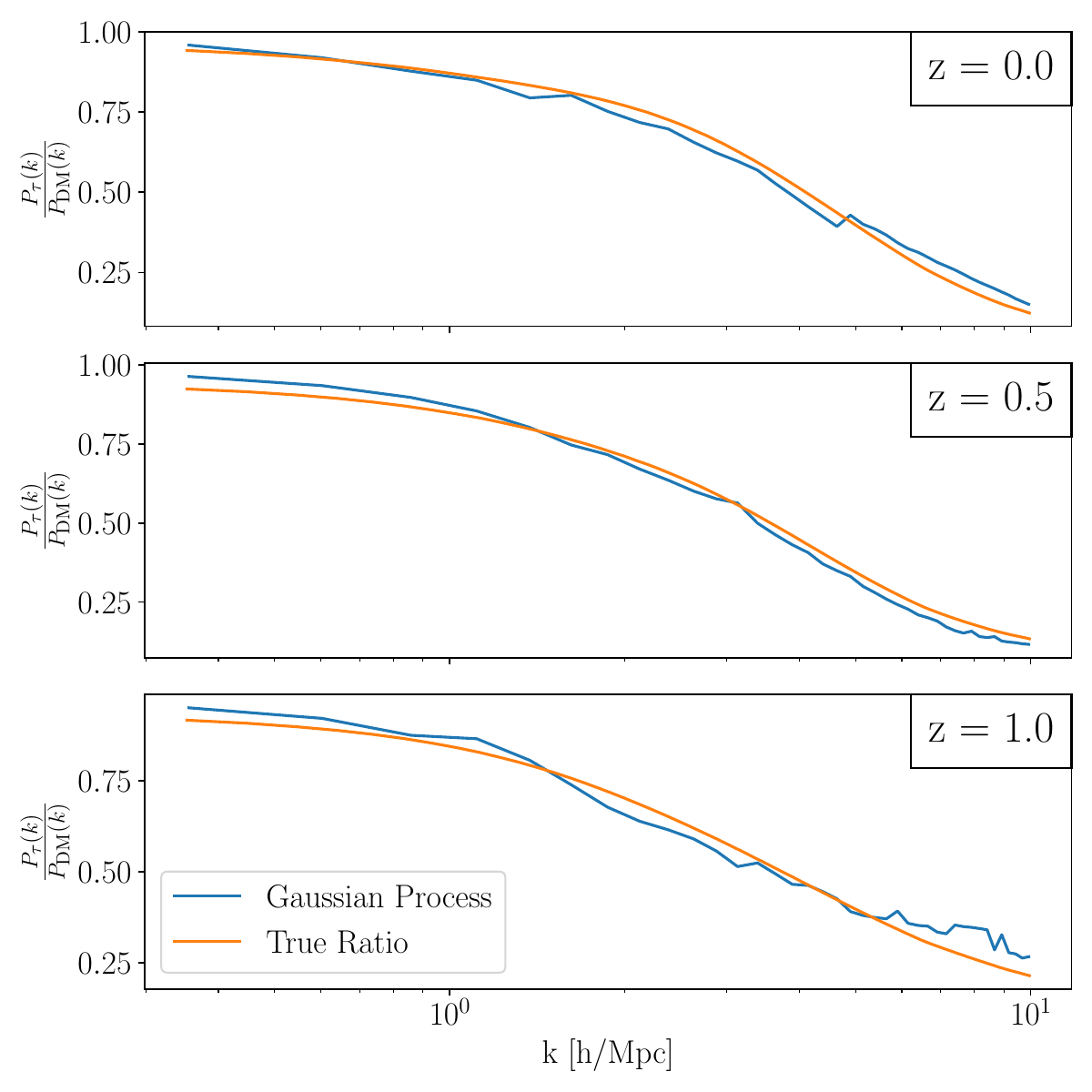}
    \caption{Real and Gaussian Process emulated power spectrum ratios for MillenniumTNG at redshifts of $0, 0.5$ and $1$. The Gaussian Process emulator is not as smooth but yields good emulation results through most of the range. The percentage error of the Gaussian Process emulator is greatest at high $k$, yielding a maximum percent error of $20\%$ at $k=10\ h/$Mpc.}
    \label{fig:GP_emulator_MTNG}
\end{figure}

Our first step in reconstructing the $\tau$ field is to test our ability to reconstruct the MillenniumTNG transfer functions using a Gaussian Process emulator trained on CAMELS. Here, we fit for the astrophysical parameters in the Gaussian Process with the true power spectrum ratio. 

The end result Gaussian Process emulators are shown in Figure \ref{fig:GP_emulator_MTNG}. We see that the Gaussian Process emulator is capable of reproducing the power spectrum ratio for MillenniumTNG to sufficient accuracy across nearly the whole range of transfer functions. However, we note differences in the emulator from the true transfer function at the higher and lower ends of the $k$ range.
To generate the transfer function using the emulator, we fix the cosmological parameters $\Omega_m$ and $\sigma_8$ to the values from MillenniumTNG, while treating the other four astrophysical parameters in the CAMELS simulations as nuisance parameters to optimize over, as discussed above in Section \ref{sec:transfer_function_emulator}.

\subsubsection{Combined reconstruction of the optical depth}
\label{sec:results_tau}

\begin{figure}
    \centering
    \includegraphics[width=\columnwidth]{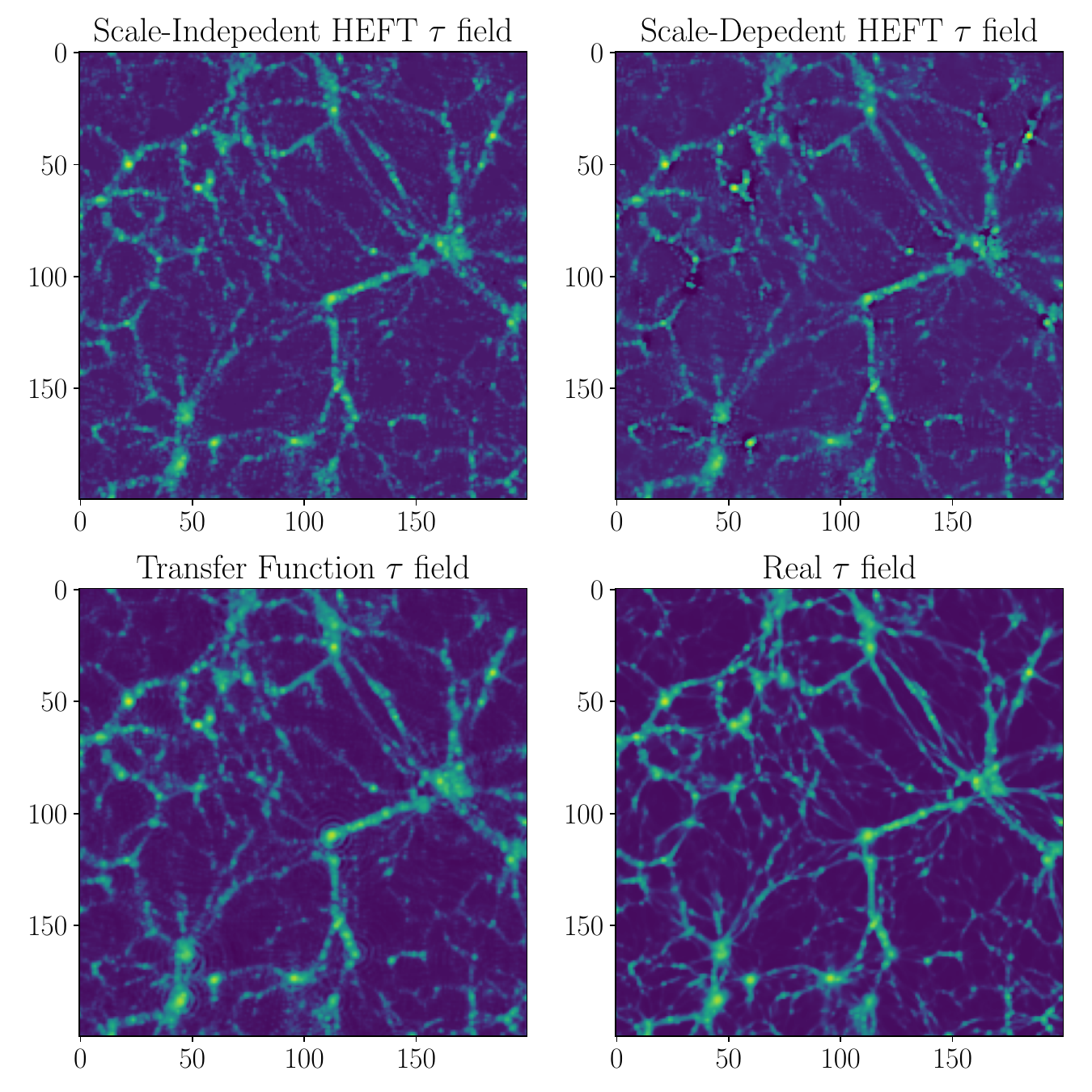}
    \caption{Cutout of a 2D slice (with a fixed index in the $z$ direction) of emulated $\tau$ fields from the scale-independent HEFT, scale-dependent HEFT, and the transfer function emulator method compared to the real $\tau$ field. We note that as described in the text, the MillenniumTNG field is pasted onto a $1080^3$ grid, so the $256\times256$ slices here represent a small section. 
    }
    \label{fig:visualization_tau}
\end{figure}

\begin{figure*}
    \centering
    \includegraphics[width=\linewidth]{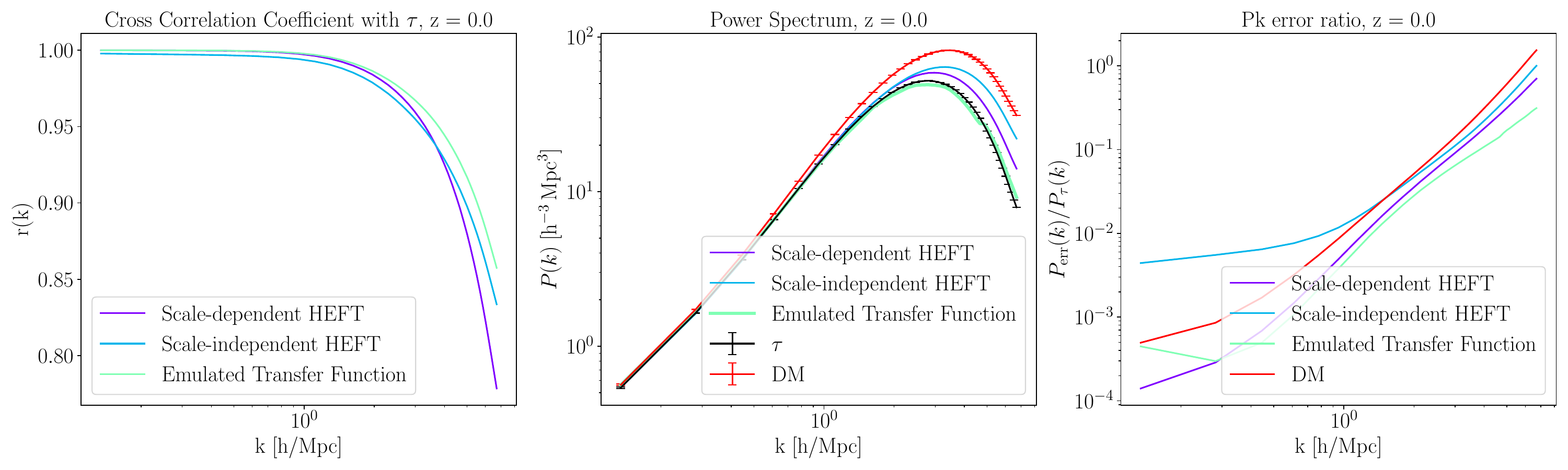}
    \caption{Correlation coefficients, power spectra and the power spectra error ratio of field reconstructions for the $\tau$ field at $z=0$. We note that as discussed in previous sections, the dark matter field has the same $r(k)$ as the field reconstructed using the transfer function emulator. We observe that the HEFT methods with worse cross-correlation and power spectrum performance than the transfer function-based method. All plots are plotted up to the Nyquist frequency cutoff.}
    \label{fig:output_plots_tau}
\end{figure*}

\begin{figure*}
    \centering
    \includegraphics[width=\linewidth]{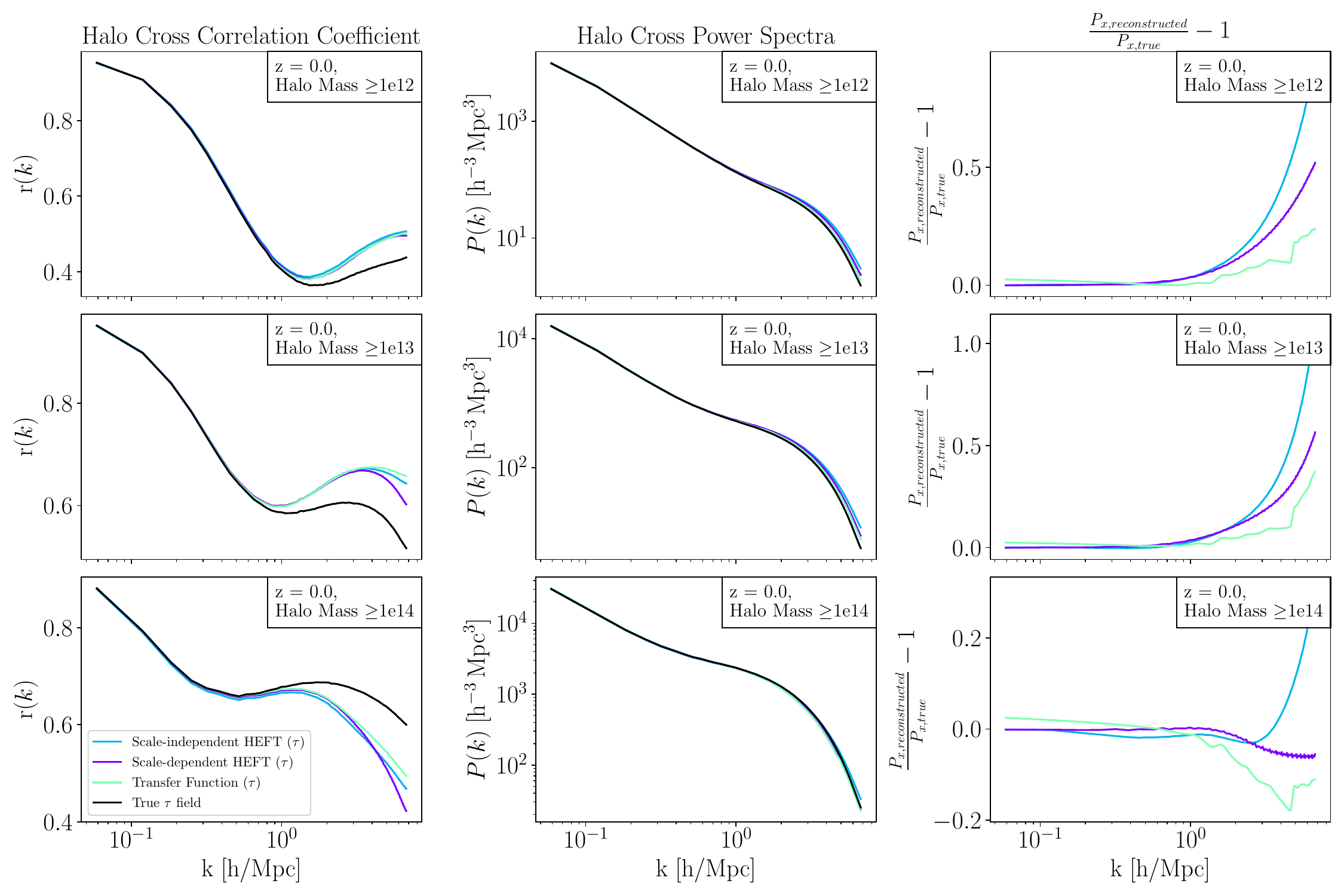}
    \caption{Halo cross correlations with the reconstructed and true $\delta_\tau$ fields, 
    cross power spectra, and cross spectra ratios of the $\tau$ field reconstructions. The reconstructed fields (especially the transfer function emulator)
    yield low cross spectra error which though increases significantly near $k=1\ h$/Mpc.}
    \label{fig:halo_plots_tau}
\end{figure*}

In testing our comparisons of the HEFT and Gaussian Process transfer function methods, we check a number of different metrics in comparing the reconstructed and real $\tau$ fields. First, we directly compare the two point statistics of the reconstructed fields with the real simulated $\tau$ field. This consists of computing the cross-correlation coefficients, the power spectrum as well as the error power spectrum, defined as 
\begin{equation}
\begin{split}
    P_{\mathrm{err}} &= \langle (\delta_{\tau} - \Tilde{\delta}_{\tau}) (\delta_{\tau} - \Tilde{\delta}_{\tau})^* \rangle \\
    &= P_{\tau, \mathrm{true}} - 2 P_{\rm cross} + P_{\tau, \mathrm{rec}}
\end{split}
\end{equation}
where $\delta_{\tau}$ is the true optical depth and $\Tilde{\delta}_{\tau}$ is the reconstructed optical depth.

Since the two-point statistics do not fully describe the reconstructed fields (and in the case of the transfer function method, the correlation coefficient with the initial dark matter field is not changed), we consider other metrics of comparison in addition to the direct comparisons described above.
As additional metrics, we also compare the two point statistics of the real and reconstructed $\tau$ fields to halo fields from MillenniumTNG, with different mass cuts. The optical depth field has important correlations with halo fields, as the profiles of the observed gas distribution around halos are an important tracer of baryon feedback. Therefore, it is important for the reconstructed optical depth to have the correct correlation with the halo field. As a result, we also check that our reconstructed fields can match the cross correlation and cross spectra statistics of the true field with haloes in MillenniumTNG, and we do this with various halo mass thresholds in the simulation.


The results for $z=0$ are shown in Figures \ref{fig:output_plots_tau} and \ref{fig:halo_plots_tau}, while results for other simulation redshifts ($z=0.5,\,1$) are similar so not shown. In these reconstructions, we see that the correlation coefficients of the scale-independent and scale-dependent HEFT reconstructions do not quite match up to the transfer function method. 
Notably, as the transfer function method by definition does not improve the cross correlation, this means that HEFT 
methods here actually perform worse than the initial dark matter simulations for the purposes of optimizing the cross correlation coefficients, especially at smaller scales. However, we see that the Gaussian Process emulated transfer function is able to replicate the power spectrum of the $\tau$ field up up to the Nyquist frequency of $k=6.78\; h/\mathrm{Mpc}$, while also producing an order of magnitude lower power spectrum error compared to the Lagrangian methods. 

In the realm of halo cross-correlations, we also see that the emulator transfer function method performs best. However, here for all mass cuts, we see a notable degradation in cross-correlations at around $k=1\; h/\mathrm{Mpc}$ from all transformed $\tau$ fields with the real $\tau$ field.

Here, we should stress that in the MillenniumTNG simulations, the $\tau$ field has a high correlation with the underlying dark matter distribution. This is not always true in other simulations with more extreme feedback, and recent work with kSZ measurements has shown that the measured gas feedback model is stronger than that given by MillenniumTNG \citep{Hadzhiyska2024_kSZ, RiedGuachalla2025}.

\subsubsection{Extended HEFT results}
\label{sec:extended_HEFT_results_tau}

 \begin{figure*}
    \centering
    \includegraphics[width=\linewidth]{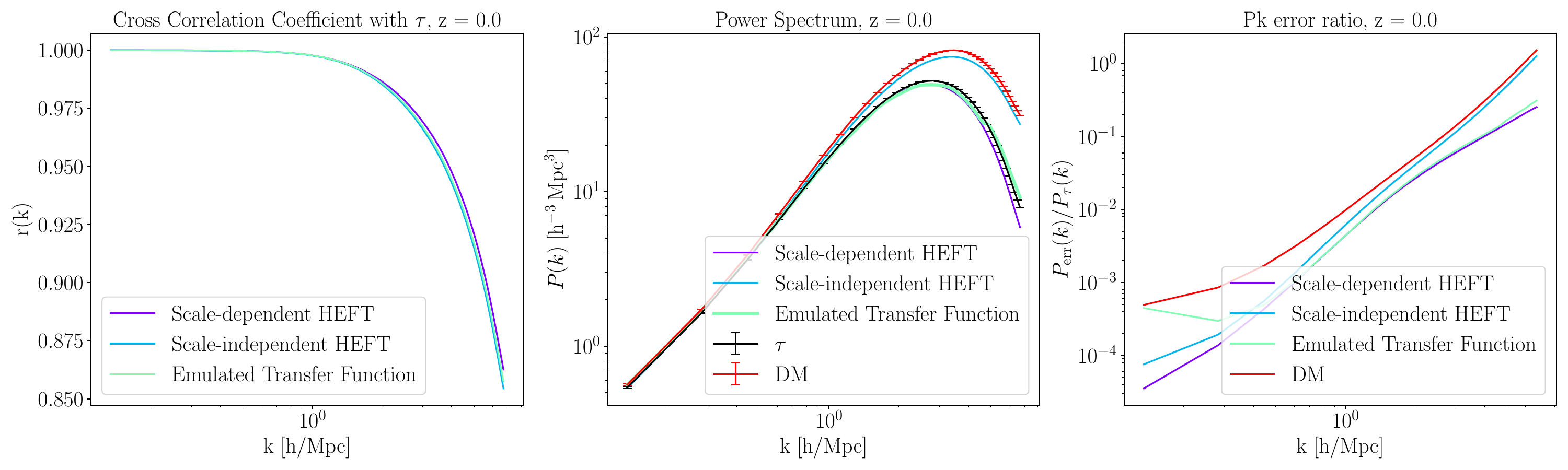}
    \caption{Extended HEFT Expansion reconstruction results on the $\tau$ field. Observe that in this case, the scale-dependent variant is able to perform as well as the transfer function method. The transfer function plots here are the same as those in Figure \ref{fig:output_plots_tau}, for ease of comparison.}
    \label{fig:extended_HEFT_tau}
\end{figure*}

Following the discussions in Section \ref{sec:extended_HEFT}, we also present the field reconstruction results for reconstructed fields using the E-HEFT formalism. Figure \ref{fig:extended_HEFT_tau} shows the correlation coefficient and power spectrum statistics of the reconstructed optical depth field using the Extended HEFT methodology. We note the improved performance of the E-HEFT methods as shown here, being able to improve slightly on the cross correlation as well as the power spectrum.

\subsection{Compton-$y$ Field}
\label{sec:resuls_Ycomp}

\begin{figure*}
    \centering
    \includegraphics[width=\linewidth]{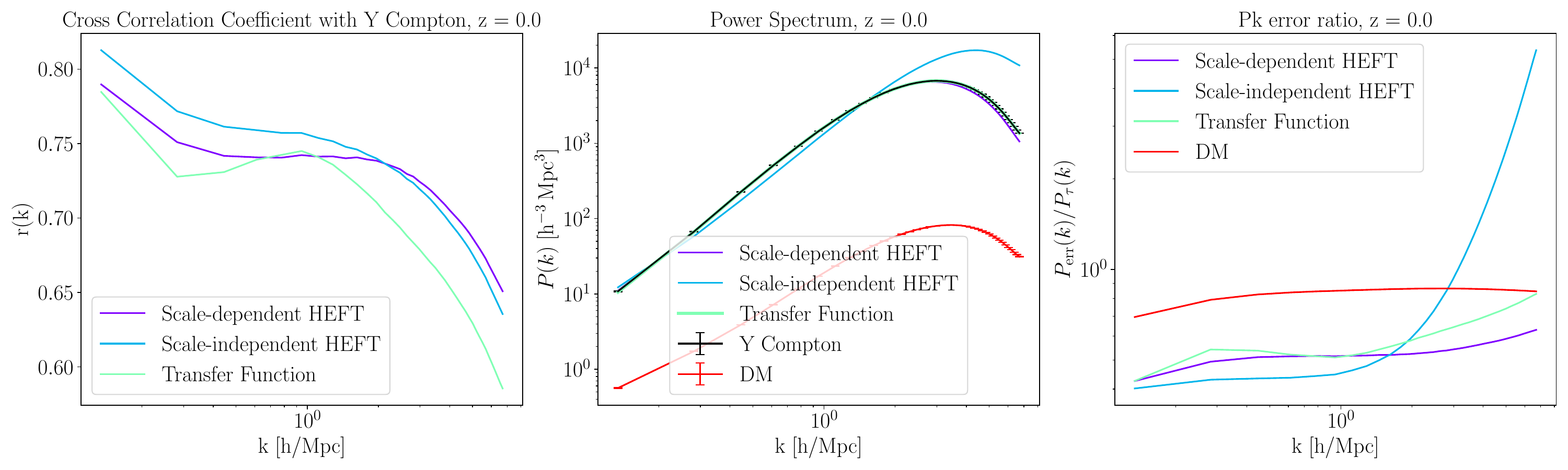}
    \caption{Cross correlation coefficients of the reconstructed fields with $\delta_{y, \rm{true}}$, 
    the power spectra and the power spectra error ratio of field reconstructions for the Compton-$y$ field. The dark matter field has the same $r(k)$ as the transfer function transformed field. We note the improved performance of the HEFT methods in this case when compared to the transfer function method across all 3 metrics shown.}
    \label{fig:output_plots_Ycomp}
\end{figure*}

\begin{figure*}
    \centering
    \includegraphics[width=\linewidth]{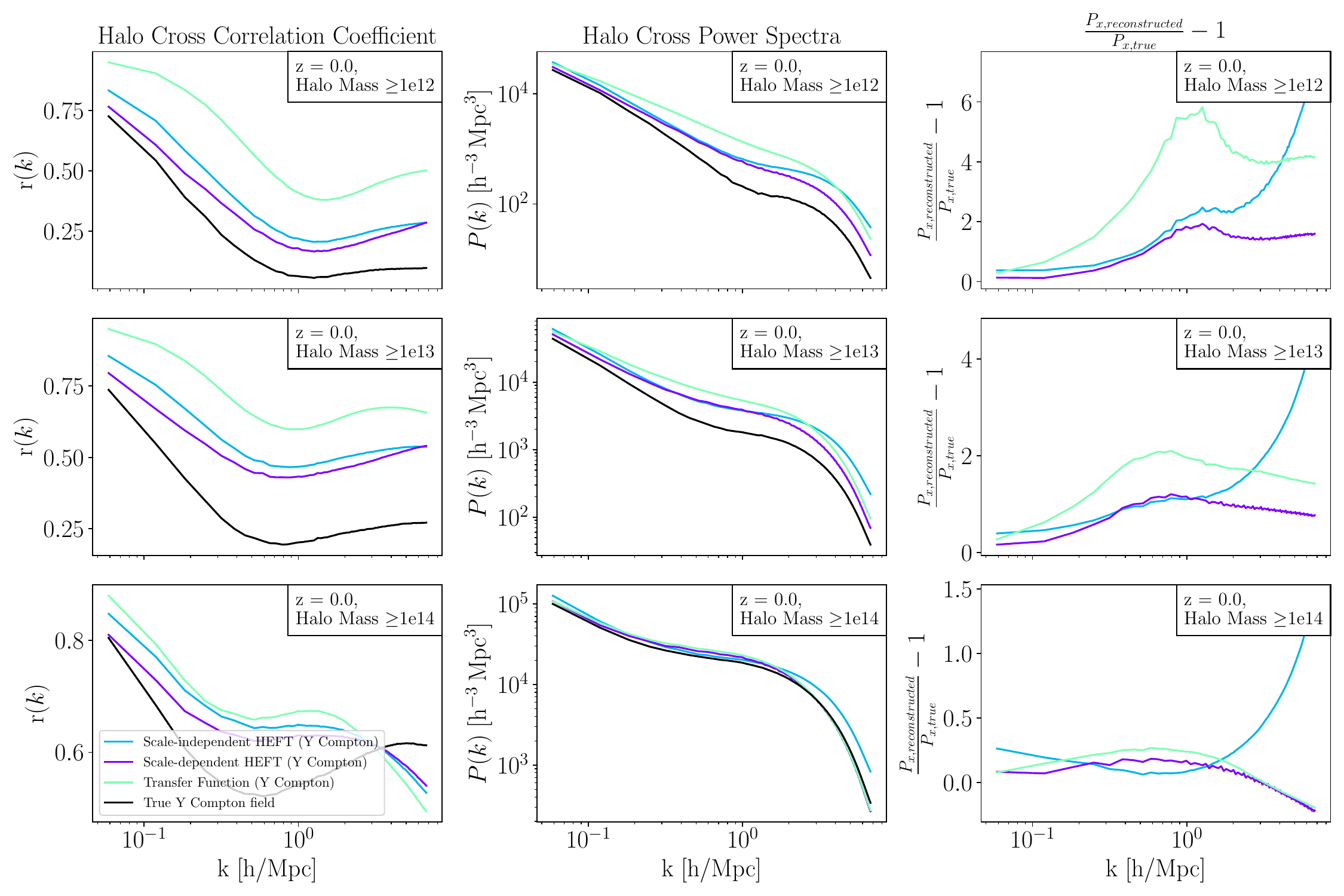}
    \caption{Halo cross correlations, cross power spectra, and cross spectra ratios of the Compton-$y$ reconstructions. We see here that the halo cross correlations are much further from truth than in the optical depth case. Future improvements could be made by re-constructing the Compton-$y$ field directly from the halo field, as there is high correlation between the two fields, especially for high halo mass thresholds.}
    \label{fig:halo_plots_Ycomp}
\end{figure*}


Another often-measured quantity related to the gas thermal pressure is the Compton-$y$ parameter, typically measured through the tSZ effect \cite{Carlstrom2002, SchaanFerraro2021, Amodeo2021, RHLiu2025}. 
In this section, we will consider the Compton-$y$ field from MillenniumTNG.
The tSZ effect is caused by the inverse Compton scattering of CMB photons interacting with free electrons of the hot ionized intergalactic gas and produces a frequency-dependent fluctuation in the CMB temperature given by: 
\begin{equation}
    \frac{\Delta T_{\mathrm{tSZ}}(\hat{\mathbf{n}})}{T_\mathrm{CMB}} = f_{\mathrm{tSZ}}(\nu)y(\hat{\mathbf{n}}) \;.
\end{equation}
Where $f_\mathrm{tSZ}$ is the frequency dependence given by
\begin{equation}
\label{eq:tSZ_SED}
   f_\mathrm{tSZ} = x \coth(x/2) - 4 \;,
\end{equation}
with $x=h\nu/k_BT_\mathrm{CMB}$, while the amplitude is given by the (dimensionless) Compton-$y$ parameter, our parameter of interest:
\begin{equation}
    y(\hat{\mathbf{n}}) = \frac{k_B\sigma_T}{m_e c^2} \int \frac{d\chi }{1+z} n_e(\chi \hat{\mathbf{n}}, z)T_e(\chi \hat{\mathbf{n}}) \;.
\end{equation}
Here, $\sigma_T$ is the Thompson cross section, $n_e$ as the free electron number density, $\chi$ as the comoving distance to redshift $z$, $m_e$ as the electron mass and $k_B$ as the Boltzmann constant.

Using the MillenniumTNG simulation suite, we can also reconstruct the Compton-$y$ parameter field in the simulation.
The process for computing the Compton-$y$ field is similar to the process of reconstruction for the optical depth field described in Section \ref{sec:optical_depth}. In this case, we compute the volume-weighted thermal pressure $P_e$ for each gas particle $i$, which is given by
\begin{equation}
    V_iP_{e,i} = x_i m_i \epsilon_i \frac{4 X_H (\gamma - 1)}{1 + 3 X_H+4X_Hx_i}
\end{equation}
where as in Section \ref{sec:optical_depth}, $x$ is the electron abundance, $m$ the mass, $\epsilon$ the internal energy, $X_H=0.76$ the primordial hydrogen mass fraction, and $\gamma=5/3$ is the adiabatic index.
The 3D map is then binned to compute the Compton-$y$ field on a $1024^3$ grid. The Compton-$y$ parameter in each cell is given by 
\begin{equation}
    y_j = \sigma_T V_j^{-1} \sum_{i \in V_j} V_i P_{{\rm e},i} ,
\end{equation}
with the same notation as in Equation \ref{eq:tau_cell}.

We note that as the CAMELS training set does not include the Compton-$y$ field, in this case, it is not possible to train a Gaussian Process emulator for the transfer function due to the lack of a training set. To substitute for this, we consider the ``best-case scenario'' (as in, the case where a Gaussian Process emulator is able to perfectly emulate the true transfer function) and use the true transfer function from the DM field to Compton-$y$ computed on the MillenniumTNG fields. 

The resulting correlation metrics are shown in Figures \ref{fig:output_plots_Ycomp} and \ref{fig:halo_plots_Ycomp}. As the correlation coefficient of the Compton-$y$ field with dark matter is lower, we see that the performance of the transfer function method is worse in this case. This is expected as the transfer function is not able to increase the cross correlation between fields, and although it is still able to replicate the two point statistics of the power spectrum in the middle panel of Figure \ref{fig:output_plots_Ycomp}, the inability of the transfer function in altering correlation of uncorrelated fields means that it does not yield a good representation for the Compton-$y$ field.

However, we see here that HEFT is able to perform much better in this case. The HEFT fit is able to increase the cross-correlation coefficient, while still matching the power spectrum using the amplitude parameter fit as discussed in Equation \ref{eq:amplitude}. Notably, the scale-independent HEFT methodology is able to improve the cross-correlation and match the power spectrum up to small scales by fitting only 4 bias parameters.


\subsubsection{Extended HEFT Results}
\label{sec:extended_HEFT_results_ComptonY}

\begin{figure*}
    \centering
    \includegraphics[width=\linewidth]{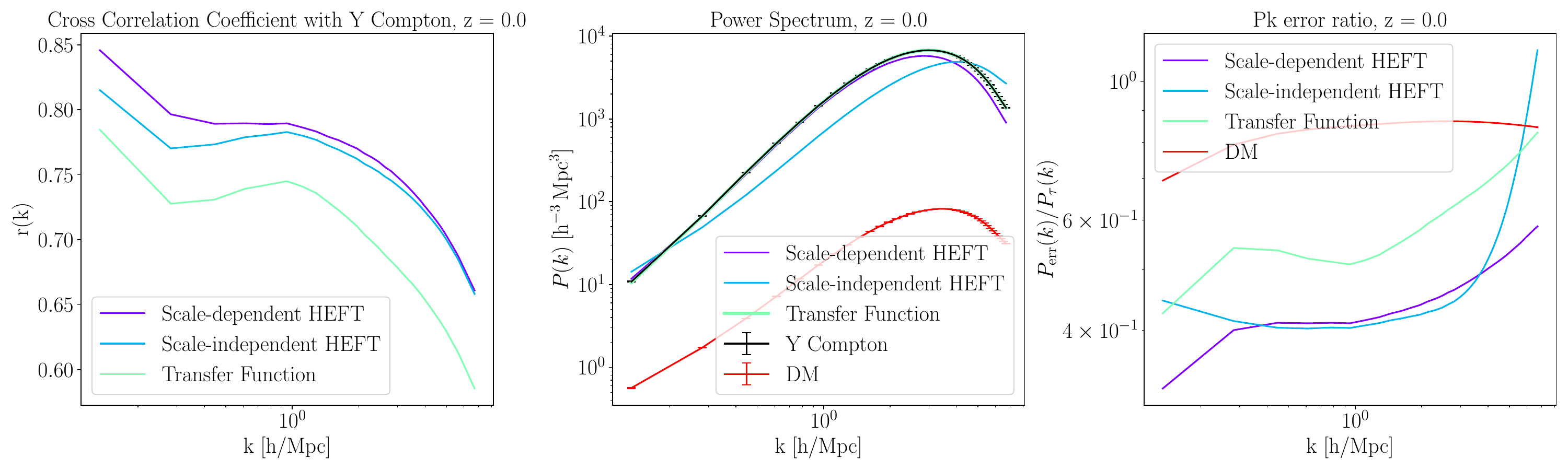}
    \caption{Extended HEFT Expansion reconstruction results on the Compton-$y$ field. Compared to Figure \ref{fig:output_plots_Ycomp}, the Compton-$y$ cross-correlation is further improved here. The transfer function plots here are the same as those in Figure \ref{fig:output_plots_Ycomp}, for ease of comparison.}
    \label{fig:extended_HEFT_YComp}
\end{figure*}

As with the optical depth reconstructions in the previous section, we also present the field reconstruction results using the E-HEFT formalism following Section \ref{sec:extended_HEFT}. The correlation coefficient and power spectrum statistics in this case are shown in Figure \ref{fig:extended_HEFT_YComp}. In this case, we can improve the performance significantly when compared to previous results in Figure \ref{fig:output_plots_Ycomp}. As the fields are less correlated to begin with, this represents an important improvement to the cross-correlation of the fields. In the case of the scale-independent E-HEFT method, we further see that 5 total bias parameters are capable of yielding a large improvement to cross correlation up to the Nyquist frequency of the box.


\section{Discussion and Conclusion}
\label{sec:discussion}

This work compared two methods for reconstructing the optical depth and Compton-$y$ parameter fields in the MillenniumTNG simulations and assessed the performance of these two methods across a number of metrics (most notably the correlation with halo fields). By interpreting the transfer function and perturbative methods, we were able to demonstrate the ability to reconstruct physically interpretable 3-dimensional 
simulation fields with high-mesh resolutions, all with relatively high computational efficiency.

The Gaussian Process (GP) -based transfer function emulator is shown to be able to effectively replicate the correct transfer function from the dark matter $\delta_{m}$ to the optical depth $\delta_{\tau}$ and Compton-$y$ $\delta_{y}$ fields. In cases where the fields were already highly correlated (i.e., $r \approx 1$), such as the optical depth field $\delta_{\tau}$, the transfer function emulator is able to accurately reproduce the target field's power spectra while retaining the already high cross correlation coefficients. However, as the transfer function method is by definition unable to increase the cross correlation coefficients between fields, we see that in cases of fields with low initial correlation (such as the $\delta_{m}$ field with the Compton-$y$ field, shown in Section \ref{sec:resuls_Ycomp}), the method does not perform optimally. 

In contrast, the Hybrid Effective Field Theory (HEFT) method showed a different set of strengths. For highly correlated fields, the field-level cross-correlation coefficients underperformed compared to the transfer function emulator method. We see a drop-off in the cross-correlation coefficient at small scales and a mismatched power spectra. However, for fields less strongly correlated with $\delta_m$ ($r \sim 0.5$) such as the Compton-$y$ field, the HEFT method improved cross-correlation coefficients by $10-15\%$. Additionally, amplitude fits described in Section \ref{sec:lagrangian_HEFT} also captured the power spectrum statistics effectively. We further note that the scale-independent HEFT approach was able to provide an efficient alternative by fitting only four free parameters, thus avoiding the need for $k$-dependent transfer functions or bias parameters.

When examining our halo cross-correlation results, we saw that the cross-correlation coefficients of the reconstructed $\delta_\tau$ field with the halo field matched the true cross-correlation coefficients
up to $k=1\ h$/Mpc, but the disagreement increased at higher $k$ values. The reconstructed Compton-$y$ fields exhibited greater discrepancies, but we still see the scale-dependent Hybrid Effective Field Theory method yielding the closest results to the true field. 
However, Figure \ref{fig:halo_plots_Ycomp} shows that the cross-correlation coefficients of the reconstructed Compton-$y$ field with the halo field is still quite different than the truth.
This, together with the results from Figure \ref{fig:output_plots_Ycomp} implies that the dark matter field is not sufficiently strongly correlated with the Compton-$y$ field. A potential avenue for improvement lies in directly reconstructing the Compton-$y$ field from the halo field or with a combination of the matter and halo fields, which is deferred to future work.

A synthesis of both methods is achieved in the Extended HEFT methodology as described in Section \ref{sec:extended_HEFT}. By introducing an additional bias parameter on the Eulerian $\delta_m$ field (analogous to the advected 1cb field in Lagrangian PT), we are able to outperform both the transfer function method and the HEFT methods using traditional Lagrangian perturbation theory.
Figures \ref{fig:extended_HEFT_tau} and \ref{fig:extended_HEFT_YComp} illustrate that the Extended HEFT bias fits match the transfer function results for highly correlated fields while further improving performance for less correlated fields like the Compton-$y$ field.

Both the transfer function emulator and HEFT methods could benefit from integration with ML techniques to enhance small-scale correlations. Recent work demonstrates the potential of diffusion models trained on CAMELS to correlate density fields effectively at small scales \cite{Ono2024}. Combining these ML methods with the large-scale performance of transfer function and EFT approaches could yield a hybrid method that retains physical accuracy on large scales while leveraging generative ML models for improvements at high $k$.

To conclude, both of the methods presented here show significant promise for reconstructing the optical depth and Compton-$y$ fields in the MillenniumTNG simulation suite. The GP-based transfer function emulator is more effective for highly correlated fields, while the HEFT approach excels in fields with less strong initial correlation. Future work could expand upon these findings by incorporating generative models to address small-scale correlations and further improve the accuracy of reconstructed simulation fields.

\acknowledgments 
We thank Shivam Pandey, Divij Sharma, Uro\v{s} Seljak, and the members of the Simons Collaboration on ``Learning the Universe'' as well as the MillenniumTNG team for many useful conversations and suggestions. 

R.H.L. is supported by the Postgraduate-Doctoral Scholarship from the Natural Sciences and Engineering Research Council of Canada (NSERC), funding reference number PGSD-567923-2022. 
B.H. is supported by the Miller Institute for Basic Science.
S.F. is supported by Lawrence Berkeley National Laboratory and the Director, Office of Science, Office of High Energy Physics of the U.S. Department of Energy under Contract No.\ DE-AC02-05CH11231.
\appendix

\bibliographystyle{prsty.bst}
\bibliography{refs}

\end{document}